# A thermodynamic and analytical description on the quantitative phase-field model with enhanced interface diffusivity

Yue Li, Lei Wang*, Junjie Li, Jincheng Wang, Zhijun Wang*

Email: lei.wang@nwpu.edu.cn

zhjwang@nwpu.edu.cn

State Key Laboratory of Solidification Processing, Northwestern Polytechnical University,

Xi'an 710072, China

## Abstract

Based on the idea of maintaining physical diffuse interface kinetics, enhancing interfacial diffusivity has recently provided a new direction for quantitative phase-field simulation at microstructural length and time scale. Establishing a general relationship between interface diffusivity and width is vital to facilitate the practical application. However, it is still limited by time-consuming numerical corrections, and its relationship with non-dilute thermodynamic properties still needs to be revealed. In this study, we present a new thermodynamic and analytical method for determining interfacial diffusivity enhancement. Unlike previous numerical corrections of partition coefficients and interface temperature, this new method aims to keep several thermodynamic quantities unchanged after enlarging the interface width. These essential quantities are theoretically proven to be diffusion potential jump across diffuse interface and free energy dissipation by trans-interface diffusion. Since no dilute approximation has been employed in model derivation, the present method is available for binary alloys with arbitrary thermodynamic properties and can be easily extended to describe multicomponent systems. Therefore, the present method is expected to advance the recent quantitative phase-field framework and facilitate its practical applications.

# 1. Introduction

In the past two decades, phase-field model (**PFM**) has been a powerful computational tool in studying material microstructure evolution kinetics [1,2]. In general, phase-field computations on a microstructural length scale require artificial interface width that is significantly larger than physical interface width. However, this will cause three spurious interface effects [3,4]: anomalous solute trapping (or chemical potential jump), interface stretching, and surface diffusion. Suppressing these artificial effects, especially spurious solute trapping, is of great significance in the quantitative phase-field simulations.

The modern quantitative phase-field model for alloys is initiated from Karma's seminal introduction of anti-trapping current in the diffusional equations [5,6], which Kim then extended to arbitrary multicomponent alloys [7]. Combining anti-trapping current with asymptotic analysis simultaneously eliminated three spurious interface effects, and phase-field kinetics were rigorously mapped to the local equilibrium conditions. Once interpolation functions are set, general anti-trapping coefficients are available for arbitrary alloys, which has been proven very successful in predicting near-equilibrium phase transformations.

Unlike near-equilibrium solidification, the thermodynamic condition at the interface is away from the local equilibrium during the rapid solidification, for which previous **PFMs** benchmarking local equilibrium fail in the quantitative description. To solve this problem, Pinomaa and Provatas [8] and Kavousi and Zaeem [9] further improved the anti-trapping method by mapping phase-field kinetics to the non-equilibrium sharp-interface model for the binary dilute alloys, i.e., the famous Continuous Growth Model (CGM) [10–12]. However, there is no simple sharp-interface description like CGM for determining the anti-trapping coefficients in (multicomponent) non-dilute alloys [13,14]. Moreover, unlike CGM available for any binary dilute alloys, sharp-interface models for non-dilute alloys differ in detail for each alloy [13,14], which means the solved anti-trapping coefficients are possibly different for each alloy. This issue will limit more practical applications, especially compared with the previous near-equilibrium anti-trapping models with general anti-trapping coefficients available for arbitrary thermodynamic properties.

Recently, Ji and Karma et al. [15,16] put forward a new quantitative **PFM** that artificially enhances interfacial diffusivity instead of using anti-trapping current and sharp-interface models. The essence of this new framework [15,16] is to maintain the realistic diffuse interface kinetics, which means that the partition coefficients and interface temperature for a physical interface width can be reproduced for a much larger width. This new framework is still under development, and a few points require further investigation. The current determination of diffusivity enhancement requires comparing numerous one-dimensional steady solutions of partition coefficients and temperature, whose numerical diffusivity also increases for (multicomponent) non-dilute alloys. Moreover, although the diffusivity enhancement was found to be independent of equilibrium partition coefficient, its relationship with other alloy thermodynamic properties is still unknown. Solving these two problems is vital to further developing the new **PFM** framework.

Converging to realistic diffuse interface kinetics is critical to Ji and Karma et al.'s new quantitative **PFM** framework [15,16]. According to classical diffuse-interface theories developed by Cahn [17] and Hillert [18,19], realistic diffuse interface kinetics are dominated by the balance between thermodynamic driving force and free energy dissipation by interface migration and trans-interface diffusion (or called solute drag). In this sense, it is more physically desirable to maintain these thermodynamic quantities after enlarging the interface width. In particular, analytical approximations for these thermodynamic quantities may exist at low- and high-velocity limits [17,18], which may simplify the determination of interfacial diffusivity enhancement. Based on this idea, this work aims to develop a thermodynamic and analytical method for determining diffusivity enhancement for various interface widths and alloy thermodynamic properties.

The structure of this article can be summarized as follows. We first show consistency between classical diffuse interface theories and phase-field kinetics. Then, using the phase-field "language", we rewrite the essential thermodynamic quantities: diffusion potential jump, thermodynamic driving force, and free energy dissipation. After this step, approximated analytical equations for determining diffusivity enhancement are reached by decoupling these thermodynamic quantities with artificial interface width. Finally, the derived interface diffusivity enhancement is validated by combining one-dimensional and two-dimensional directional rapid solidification.

## 2. A brief review of classical diffuse interface theory

The modern continuum description of diffuse interface kinetics began with J.W. Cahn's seminal solute drag theory for grain boundary motion [17]. Then, M. Hillert gave a general framework available for phase transformations [18,19]. The core of their theories is that the diffuse interface velocity is determined by the balance of total driving force on the interface and the free energy dissipation by interface migration and trans-interface diffusion.

Applying their theories in solidification and neglecting solid diffusion, the diffuse solid-liquid interface velocity ($V$) can be generally expressed as

$$\frac{1}{V}(Q_M + Q_J) = \Delta f_{tot}, \tag{1}$$

where $\Delta f_{tot}$ is the total driving force that expressed as

$$\Delta f_{tot} = f_l^* - f_s^* - \tilde{\mu}_l^*(c_l^* - c_s^*), \tag{2}$$

where $f_l^*$ ($f_s^*$) and $c_l^*$ ($c_s^*$) represent the liquid (solid) free energy and concentration at the diffuse interface boundary, and $\tilde{\mu}_l^* = \partial f_l^*/\partial c_l^*$ is the corresponding diffusion potential. As for the free energy dissipation by interface migration ($Q_M$) and trans-interface diffusion ($Q_J$), they follow the standard quadratic form in the classical irreversible thermodynamics [20,21]

$$Q_M = \frac{V^2}{m}, \tag{3}$$

$$Q_J = \frac{1}{v_m} \int_{-\eta/2}^{\eta/2} \frac{J^2}{M_c} dx, \tag{4}$$

where $m$ is the intrinsic interface mobility after incorporating the solute drag effects [14,22], $M_c$ is the atomic mobility for diffusion flux $J$, and the $\eta$ represents the interface width. The quantity $Q_J/V$ is usually called as the solute drag force [18,19].

The most important thing that Eq. (1) tells us is that $Q_M$ and $Q_J$ determine the diffuse interface kinetics once the boundary conditions in the $\Delta f_{tot}$ are fixed. In later Sec. 3.2, we will show that phase-field equations at the 1-D steady state can yield the same free energy balance as Eqs. (1)-(4). **Therefore, the key to quantitative phase-field modeling is maintaining these thermodynamic quantities after artificially enlarging the interface width.**

## 3. Phase-field model with enhanced interface diffusivity
### 3.1 Governing equations

In this work, we adopt the Kim-Kim-Suzuki (**KKS**) model [23] to describe binary alloys with arbitrary thermodynamic properties. Using the double-obstacle potential $\phi_s(1-\phi_s)$ [24], governing equations can be summarized as follows:

$$\frac{1}{M_\phi}\frac{\partial \phi_s}{\partial t} = \frac{\partial h_s}{\partial \phi_s}\frac{1}{v_m}[f_l - f_s - \tilde{\mu}(c_l - c_s)] + \frac{4\sigma}{\eta}(2\phi_s - 1) + \frac{8\eta\sigma}{\pi^2}\nabla^2 \phi_s, \tag{5}$$

$$\frac{\partial c}{\partial t} = \nabla \cdot (A_\phi M_c)\nabla\tilde{\mu}, \tag{6}$$

In Eq. (5), the quantities $f_s$ ($f_l$) and $c_s(c_l)$ are the free energy and separated concentration field of solid (liquid) phase, $v_m$ is the molar volume, $\sigma$ is the solid-liquid interface energy, $M_\phi$ is the phase-field mobility, and $\eta$ still represents the interface width. In Eq. (6), the quantity $c$ represents the total concentration, $\tilde{\mu}$ means the diffusion potential, and $M_c$ is the atomic mobility for diffusion. In the **KKS** framework [23], the separated concentration fields $c_s$ and $c_l$ are solved by combining equal diffusion potential condition with the mixture rule:

$$\tilde{\mu} = \frac{\partial f_s}{\partial c_s} = \frac{\partial f_l}{\partial c_l}, \tag{7}$$

$$c = h_s c_s + h_l c_l. \tag{8}$$

where $h_s$ and $h_l$ are interpolation functions for concentration and free energy.

Now, inspired by Ji and Karma et al.'s recent work [15,16], an enhancing function $A_\phi$ is applied in the diffusion equation [Eq. (6)] to suppress spurious solute trapping,

$$A_\phi = 1 + a\sqrt{\phi_s \phi_l}; (a \geq 1), \tag{9}$$

which is slightly different from Ji and Karma et al.'s quadratic form. Frankly, there are various choices for $A_\phi$, ensuring that the enhancement only works on the interfacial region from $\phi_s = 0$ to $\phi_s = 1$. The remaining content is aimed at establishing a relationship between enhancing parameter $a$ and interface width $\eta$.

### 3.2 One-dimensional steady analysis

Before proceeding, we use **Fig. 1** to visualize diffuse interface scenario from solid boundary ($\phi_l = 0.001$) to liquid boundary ($\phi_l = 0.999$), and the superscript * denotes the physical quantities at the boundaries of the diffuse interface.

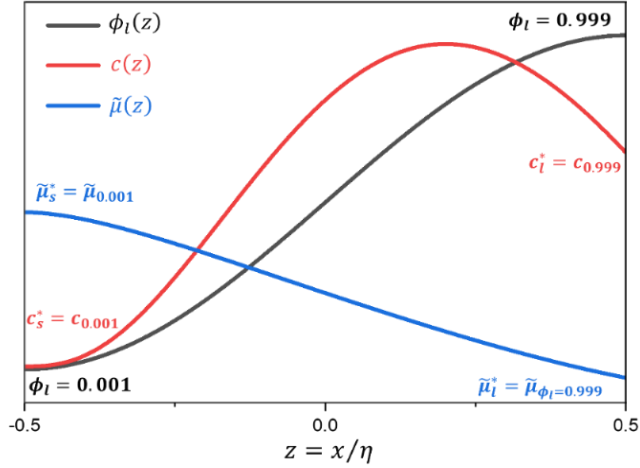

**Fig. 1.** Definitions of concentrations ($c_s^*$, $c_l^*$) and diffusion potentials ($\tilde{\mu}_s^*$, $\tilde{\mu}_l^*$) at the solid ($\phi_l = 0.001$) and liquid ($\phi_l = 0.999$) phase boundary in the diffuse interface scenario.

As the first step, let us define the jump of diffusion potential across the diffuse interface: $\Delta\tilde{\mu}^* = \tilde{\mu}_l^* - \tilde{\mu}_s^*$, which determines the boundary concentrations ($c_l^*$, $c_s^*$) and thus the total driving force [$\Delta f_{tot}$, Eq. (2)]. Applying the standard coordinate transformation $\partial/\partial t = -V \partial/\partial x$, Eq. (6) is integrated into

$$J = V(c_s^* - c) = A_\phi M_c \frac{\partial \tilde{\mu}}{\partial x}, \quad (10)$$

which, after further integration, transforms into

$$\Delta \tilde{\mu}^* = \int_{-\eta/2}^{\eta/2} \frac{V(c_s^* - c)}{A_\phi M_c} dx. \quad (11)$$

It should be mentioned that Eq. (10) is based on a precondition of sluggish diffusion flux at the solid boundary ($\phi_l = 0.001$). In other words, it still belongs to the one-side diffusion model as done in most previous quantitative **PFMs** [5–9,15].

As the second step, we can now define the balance of total driving force and different free energy dissipation by interface migration and trans-interface diffusion. Using steady diffusion flux [Eq. (10)] and applying $\partial/\partial t = -V \partial/\partial x$ again to Eq. (5), we can derive

$$\Delta f_{tot} = \frac{1}{V}(Q_{M,PF} + Q_{J,PF}) \quad (12)$$

where $\Delta f_{tot}$ was given in Eq. (2), $Q_{M,PF}$ and $Q_{J,PF}$ represent free energy dissipation of interface migration and trans-interface diffusion within the phase-field framework, which are given by

$$Q_{M,PF} = \frac{V^2}{M_\phi} \frac{\pi}{8\eta}, \quad (13)$$

$$Q_{J,PF} = \frac{V^2}{v_m} \int_{-\eta/2}^{\eta/2} \frac{(c_s^* - c)^2}{A_\phi M_c} dx. \quad (14)$$

Comparing Eqs. (12)-(14) with Eqs. (1)-(4), we can find phase-field model yield the same balance between driving force and dissipation as the classical diffuse interface theories developed by Cahn [17] and Hillert [18,19]. Combining Eq. (13) with Eq. (3), the phase-field mobility $M_\phi$ and intrinsic interface mobility $m$ are related by

$$M_\phi = \frac{\pi}{8\eta} m, \quad (15)$$

which is also similar to Ji and Karma et al.'s recent work [15,16]. It should be emphasized the present interface mobility $m$ differs from Kim's previous work [7,23], whose interface mobility is defined at the near-equilibrium condition. Instead, as done in atomistic simulations [14,22], the present $m$ is usually measured after considering the solute drag effects. With $M_\phi$ defined in Eq. (15), it is obvious that $Q_{M,PF}$ [Eq. (13)] is not affected by enlarging interface width. **Therefore, the remaining task aims to make integrals in $\Delta\tilde{\mu}^*$ [Eq. (12)] and $Q_{J,PF}$ [Eq. (14)] decoupled with interface width**, which determines the total driving force ($\Delta f_{tot}$) and solute drag ($Q_J/V$), respectively.

## 4. Interfacial diffusivity Enhancement

Because $\Delta\tilde{\mu}^*$ [Eq. (11)] and $Q_{J,PF}$ [Eq. (14)] are related to the concentration field $c$ and thus depending on the interface velocity $V$, it is difficult to find analytical expressions for arbitrary $V$. However, we can seek some approximations at the low- and high-velocity regions. In the following subsections, we first derive the enhancement of interface diffusivity $A_\phi$ [Eq. (9)] by decoupling $\Delta\tilde{\mu}^*$ [Eq. (11)] and $Q_{J,PF}$ [Eq. (14)] with interface width ($\eta$) at the low- and high-velocity regions. In particular, with a proper selection of atomic mobility $M_c$ [Eq. (6)], it is hoped to find $A_\phi$ independent of the material systems, which is convenient for practical applications like the previous anti-trapping models for near-equilibrium solidification. In later **Sec. 5**, we then discuss the possibility of extending $A_\phi$ (for low- and high-velocity condition) to more interface velocities.

### 4.1 Low-velocity limit

At the low-velocity regime where interface Pecklet number $P = \eta V/2D_L \ll 1$, the total concentration profile $c(x)$ is almost independent of interface velocity, which can be approximated by the equilibrium value

$$\begin{aligned}c_s(x) = c_s^{eq}; c_l(x) = c_l^{eq}\\ c(x) \approx c^{eq} = h_s c_s^{eq} + h_l c_l^{eq}\end{aligned} \quad (16)$$

which has also been derived from the previous asymptotic analysis [5,6,8,9] and applied to derive the anti-trapping coefficients [7]. To intuitively see this approximation, **Fig. 2** presents an example of Si-9at.%As alloy with a small Pecklet number $P = 0.01$. All thermodynamic and kinetic parameters are the same as used in the following **Sec. 5**.

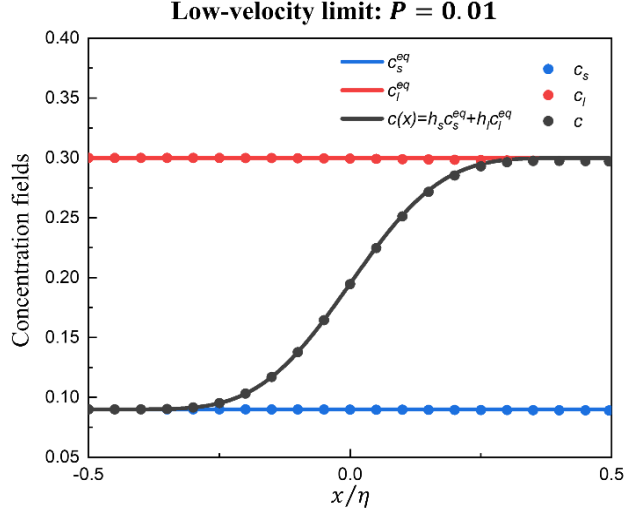

**Fig. 2.** Concentration profiles across the diffuse interface at the low-velocity limit, where the numerically solved concentration fields $[c_s(x), c_l(x)]$ are close to equilibrium values $[c_s^{eq}, c_l^{eq}]$.

To make $M_c$ in Eqs. (11) and (14) decoupled with their integrals, we assume $M_c$ at low-velocity regime follow the mixture rule in the previous **KKS** model [7]:

$$M_c^{low-V} = h_l M_l, \tag{17}$$

where $M_l = D_l(\partial^2 f_l / \partial c_l^2)^{-1}$ is the atomic mobility for liquid diffusion and the solid contribution was ignored for solidification ($M_s = 0$). We further substitute Eqs. (16) and (17) into the diffusion potential jump $\Delta\tilde{\mu}^*$[Eq. (11)] and trans-interface diffusion dissipation $Q_{J,PF}$ [Eq. (14)], yielding

$$\Delta\tilde{\mu}^*_{low-V} = \frac{V(c_s^{eq} - c_l^{eq})\eta}{M_l^{eq}}\frac{1}{\pi}\int_0^1 \frac{1}{A_\phi\sqrt{\phi_s\phi_l}}d\phi_s, \tag{18}$$

and

$$\begin{aligned} Q_{J,PF}^{low-V} &= \frac{V^2(c_s^{eq} - c_l^{eq})^2}{M_l^{eq} v_m}\frac{\eta}{\pi}\int_0^1 \frac{h_l}{A_\phi\sqrt{\phi_s\phi_l}}d\phi_s. \\ &\approx \frac{V^2(c_s^{eq} - c_l^{eq})^2}{M_l^{eq} v_m}\frac{\eta}{\pi}\frac{1}{2}\int_0^1 \frac{1}{A_\phi\sqrt{\phi_s\phi_l}}d\phi_s. \end{aligned} \tag{19}$$

One can examine that the degree of approximation in Eq. (19) is very high, no matter what the value of $a$ in Eq. (9). It should be mentioned that the gradient of phase-field parameter was replaced by its equilibrium value $[\partial\phi_s/\partial x = -(\pi/\eta)\sqrt{\phi_s\phi_l}]$, during the derivations of Eqs. (18) and (19).

Now, one can see that $\Delta\tilde{\mu}^*_{low-V}$ and $Q^{low-V}_{J,PF}$ are simultaneously determined by the integral $\eta \int 1/A_\phi\sqrt{\phi_s\phi_l}\,d\phi_s$. At the low-velocity regime, if the diffuse interface kinetics are not changed by enlarging $\eta$, the enhancing parameter $A_\phi$ should maintain the integral

$$\int_0^1 \frac{1}{(1+a\sqrt{\phi_s\phi_l})\sqrt{\phi_s\phi_l}}d\phi_s = \pi\frac{\eta_0}{\eta}. \tag{20}$$

where $\eta_0$ is the physical interface width and $\eta$ here represents the artificial interface width used in the microstructure simulations.

### 4.2 High-velocity limit

At the high-velocity regime where interface Peclet number $P = \eta V/2D_L \gg 1$, there is no significant diffusion boundary within the bulk region ($\phi_l=1$). In other words, both concentrations at the boundaries of diffuse interface are close to the nominal concentration: $c_s^* = c_l^* = c_0$. Then, the diffusion potentials at the phase boundaries are $\tilde{\mu}_s^*(c_0)$ ($\phi_l = 0.001$) and $\tilde{\mu}_l^*(c_0)$ ($\phi_l = 0.999$), which are almost independent of interface velocity. In addition, since diffusion potential changes from solid side to liquid side monotonically, we may give a first order approximation of $\tilde{\mu}(x)$, yielding

$$\tilde{\mu}(x) = \tilde{\mu}_s^*(c_0) + [\tilde{\mu}_l^*(c_0) - \tilde{\mu}_s^*(c_0)]h_l. \tag{21}$$

Also, to intuitively see this approximation, **Fig. 3** gives an example of Si-9at.%As alloy with large Peclet number $P = 10$. Now, we substitute Eq. (21) into Eq. (10) and obtain that

$$\frac{V(c_0 - c)}{A_\phi M_c} = [\tilde{\mu}_l^*(c_0) - \tilde{\mu}_s^*(c_0)]\frac{\partial h_l}{\partial \phi_s}\frac{\partial \phi_s}{\partial x}. \tag{22}$$

Substituting Eq. (21) into Eqs. (11), diffusion potential jump at high-velocity regime $\Delta\tilde{\mu}^*_{high-V}$ naturally recovers a constant: $\Delta\tilde{\mu}^*_{high-V} = \tilde{\mu}_l^*(c_0) - \tilde{\mu}_s^*(c_0)$, which is expected at the beginning of this subsection. Thus, we here only focus on the trans-interface dissipation at the high-velocity ($Q_{J,PF}^{low-V}$) by substituting Eq. (22) into Eq. (14)

$$Q_{J,PF}^{high-V} = \frac{(\Delta\tilde{\mu}_{high-V}^*)^2}{v_m} \frac{\pi}{\eta} \int_0^1 A_\phi M_c \left(\frac{\partial h_l}{\partial \phi_s}\right)^2 \sqrt{\phi_s \phi_l}\, d\phi_s. \tag{23}$$

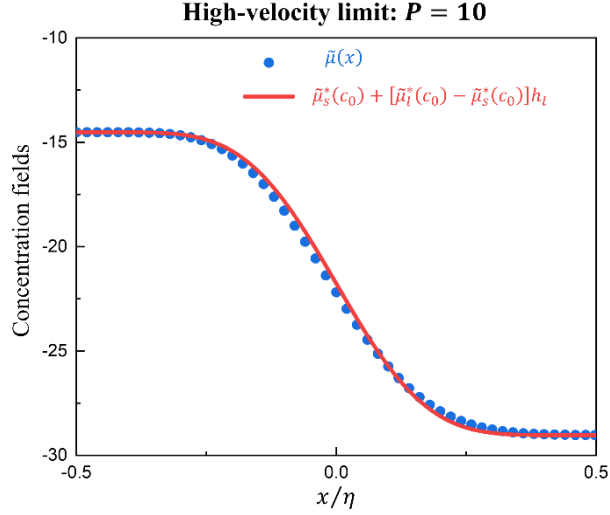

**Fig. 3.** Diffusion potential across the diffuse interface at the high-velocity limit, where the numerically solved concentration fields [$\tilde{\mu}(x)$] are close to the approximation in Eq. (21).

Now, to make $M_c$ decoupled with integral in Eq. (23), we here assume another interpolation method

$$M_c^{high-V} = h_l D_l \left(h_s \frac{\partial^2 f_s}{\partial c_s^2} + h_l \frac{\partial^2 f_l}{\partial c_l^2}\right)^{-1}, \tag{24}$$

which also ensures that $M_c^{high-V}(h_l = 1) = M_l$ and $M_c^{high-V}(h_l = 0) = 0$, i.e., recovering the classical diffusion kinetics. For the dilute alloys, one can see that Eq. (24) reproduces $M_c^{high-V} = h_l D_l c/RT$ in Karma et al.'s previous works [5,6,16]. Moreover, even for the non-dilute alloys, the interpolation in Eq. (24) can usually make $M_c^{high-V}$ is close to a function of total concentration $c$. Since that $c$ is close to nominal concentration $c_0$ at high velocities, Eq. (24) can make $Q_{J,PF}^{high-V}$ independent of material systems

$$Q_{J,PF}^{high-V} \approx \frac{(\Delta\tilde{\mu}_{high-V}^*)^2 D_l}{v_m \left(h_s \frac{\partial^2 f_s}{\partial c_s^2} + h_l \frac{\partial^2 f_l}{\partial c_l^2}\right)} \frac{\pi}{\eta} \int_0^1 A_\phi h_l \left(\frac{\partial h_l}{\partial \phi_s}\right)^2 \sqrt{\phi_s \phi_l}\, d\phi_s. \tag{25}$$

Now, if we want to maintain $Q_{J,PF}^{high-V}$ after enlarging interface width, the enhancing parameter $a$ should meet the condition

$$\int_0^1 h_l(1+a\sqrt{\phi_s\phi_l})\left(\frac{\partial h_l}{\partial \phi_s}\right)^2 \sqrt{\phi_s\phi_l}\, d\phi_s = \frac{\eta}{\eta_0}\int_0^1 h_l\left(\frac{\partial h_l}{\partial \phi_s}\right)^2 \sqrt{\phi_s\phi_l}\, d\phi_s, \quad (26)$$

where $\eta_0$ and $\eta$ still denote physical and artificial interface widths, respectively. Once interpolation function $h_{s,l}$ is set, Eq. (26) can yield elegant analytical solutions of bridging enhancing parameter $a$ and ratio of interface width $\eta/\eta_0$.

### 4.3 Comparison between low- and high-velocity solutions

Eqs. (20) and (26) provide two equations of determining the enhancing parameter $a$ in Eq. (9). Since no dilute approximation was adopted in the derivations, Eqs. (20) and (26) are expected to describe arbitrary binary alloys. Now, we are curious about which one is more accurate to apply, and which can be extended to more interface velocities.

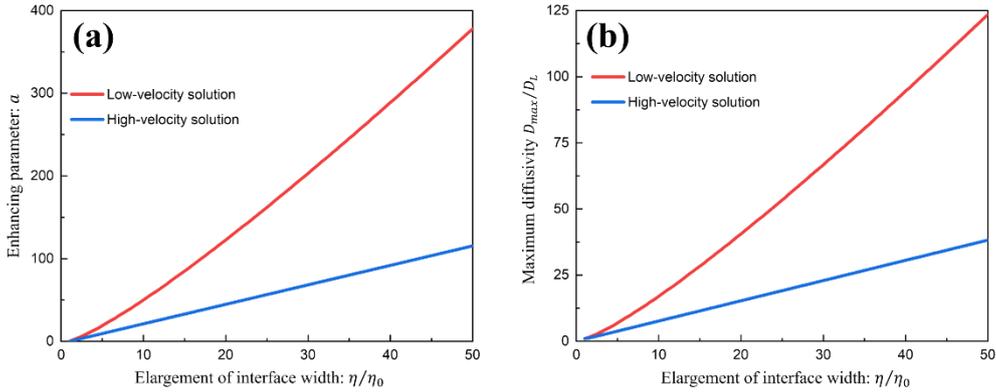

**Fig. 4.** Comparisons of (a) enhancing parameter $a$ and (b) the maximum diffusivity $D_{max}/D_L$ predicted by Eq. (20), Eq. (27), and Eq. (28).

To answer these questions, we first compare predictions of Eqs. (20) and (26). As an example, we choose interpolation function as $h_s = \phi_s$, which can yield a simple analytical solution of Eq. (26),

$$a_{high-V} = \frac{3\pi}{4}\left(\frac{\eta}{\eta_0}-1\right). \quad (27)$$

As compared in **Fig. 4(a)**, we can find $a_{low-V}$ is significantly larger $a_{high-V}$, which is not strange because larger $a$ is needed to maintain the local-equilibrium condition at the low-velocity regime. In addition, it is known that numerical stability of diffusion equation is generally proportionate to $\Delta t \propto (dx)^2/D_{max}$, and the ratio $D_{max}/D_L$ can be calculated by

$$\frac{D_{max}}{D_L} = max\{A_\phi h_l\}. \tag{28}$$

which are shown in **Fig. 4(b)**. Therefore, due to the limitation of numerical stability ($\Delta t$), it is better to use $a_{high-V}$ since its prediction of $D_{max}/D_L$ is also significantly smaller than $a_{low-V}$.

## 5. Model verification

In former Sec. 4, we have yielded two analytical solutions for low- and high-velocity regions, respectively, and briefly compared their predictions and showed that the high-velocity solution [Eq. (26)] is advantageous due to its simpler expression [e.g., Eq. (27)] and better numerical stability [Eq. (28)]. In this section, we shall test these two solutions by combining one-dimensional (1D) steady solutions and two-dimensional (2D) steady simulations. In particular, we want to explore whether these two solutions can be extended to more interface velocities, which means the low-velocity (high-velocity) solution can be applied to describe intermediate and high (low) velocities.

As usually done in theoretical works for rapid solidification kinetics, this work adopts the Si-9at%As alloy [25,26] as a comparison object. Although numerous works have used this alloy, a significant discrepancy exists regarding the liquid diffusion coefficient $D_l$. In Kittl et al.'s original work [26], they reported that $D_l = 1.5 \times 10^{-8} \ m^2/s$, and this value has been used by Pinomaa and Provatas [8] and Kavousi and Zaeem [9] for examinations of their anti-trapping models. However, if using such a diffusivity, we cannot reproduce the experimentally measured partition coefficients by using the physical interface width ($\sim nm$) in the classical diffuse interface model [18,19]. This has been also reported by Kittl et al. in their original paper [26]. Due to these reason, one can find that a much smaller diffusivity $D_l =$ $1.5 \times 10^{-9} \ m^2/s$ was usually adopted in several phase-field works with a physical interface width, e.g. Galenko et al. [27], Danilov and Neslter [28], Wang et al. [29–31], and Steinbach et al. [32]. Because this work aims to make phase-field predictions converging to the classical diffuse interface theories, we choose this smaller diffusivity $D_l = 1.5 \times 10^{-9} \ m^2/s$. Additionally, even for the thermodynamic properties, several works have reported the non-dilute behaviors of Si-9at%As alloy [13,33], whereas most theoretical works still use the dilute approximation. However, due to a lack of precise thermodynamic database for Si-9at%As alloy, we herein still adopts the dilute form by treating Si-9at%As alloy as a model system. The corresponding parameters are summarized as follows: equilibrium partition coefficient $k_e = 0.3$, melting point $T_m = 1685K$, equilibrium liquid slope $-400 \ K/at.\%$, molar volume $v_m = 1.2 \times 10^{-5} \ m^3/mol$, the solid-liquid interface energy $\sigma = 0.477 \ J/m^2$, diffusion coefficients $D_l = 1.5 \times 10^{-9} \ m^2/s$ and $D_s = 3 \times 10^{-13} \ m^2/s$, physical interface mobility $m = 2.56 \times 10^{-8} \ m^4/Js$, and capillary anisotropy strength $\varepsilon = 0.03$.

**5.1 One-dimensional steady solution**

As a first step for model verification, we solve the 1-D steady solutions without expanding the interface width, which means the enhancing function is kept as $A_\phi = 1$. Approximating $\phi(x)$ as its equilibrium profile $\phi_{eq}(x) = 0.5 - 0.5 sin(\pi x/\eta)$, the numerical procedure for solving 1-D steady solutions can be summarized as follows. Setting interface velocity $V = 0.01 - 1 \ m/s$ and assuming solid boundary concentration as $c_s^* = 0.09$, we numerically iterate Eq. (10) and Eq. (12) to obtain the steady concentration profile and the corresponding interface temperature. As for the partition coefficient $k_V$, we adopt the most widely used expression, which is the ratio between solid boundary concentration $c_s^*$ and the maximum concentration within the interface region $max\{c(x)\}$: $k_V = c_s^*/max\{c(x)\}$. During the former derivations in Sec. 4, although we have used two interpolation methods for atomic mobility $M_c$ [Eqs. (17) and (24)], the numerical results in **Fig. 5** show that there is no significant difference. When using physical interface width $\eta_0 \sim 4nm$, both $M_c$ can yield a similar $k_V - V$ relation as the experimental report.

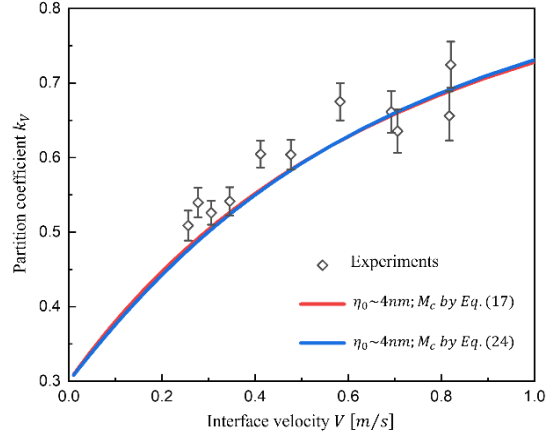

**Fig. 5.** One-dimensional steady solution for Si-9at.%As alloy with a physical interface width, in comparison with Kittl et al.'s experiments [25,26].

Now, we solve the 1-D steady solutions with the enlarged interface width. The whole interface velocity is divided into two regions: from low- to intermediate-velocity $V = 10^{-6} - 10^{-2}\ m/s$; and from intermediate- to high-velocity $V = 10^{-2} - 1\ m/s$. An average of two solutions in **Fig. 5** was assumed as a standard solution.

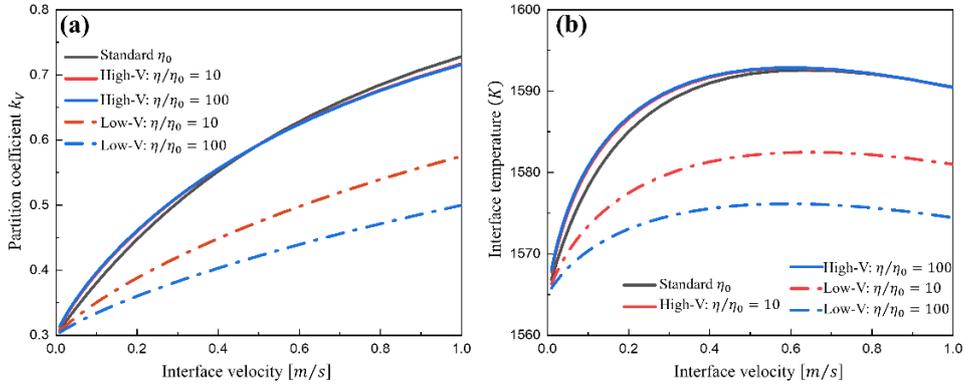

**Fig. 6.** Comparisons of 1-D steady solutions before and after enlarging the interface width from intermediate velocity $V = 10^{-2}\ m/s$ to high velocity $V = 1\ m/s$.

Let us first focus on $V = 10^{-2} - 1\ m/s$. As shown in **Fig. 6**, the high-velocity solution by Eq. (27) converges well to the standard solution. In particular, its convergence of interface temperature increases with increasing interface velocity, which is consistent with the theoretical expectation in **Sec. 4.2**. In contrast to high-velocity solution, the low-velocity solution by Eq. (20) fails in reproducing the standard solution. In this case, the physically non-equilibrium interface effects, e.g., solute trapping, are oversuppressed by the significantly larger interfacial diffusivity, as shown in previous **Fig. 4(b)**.

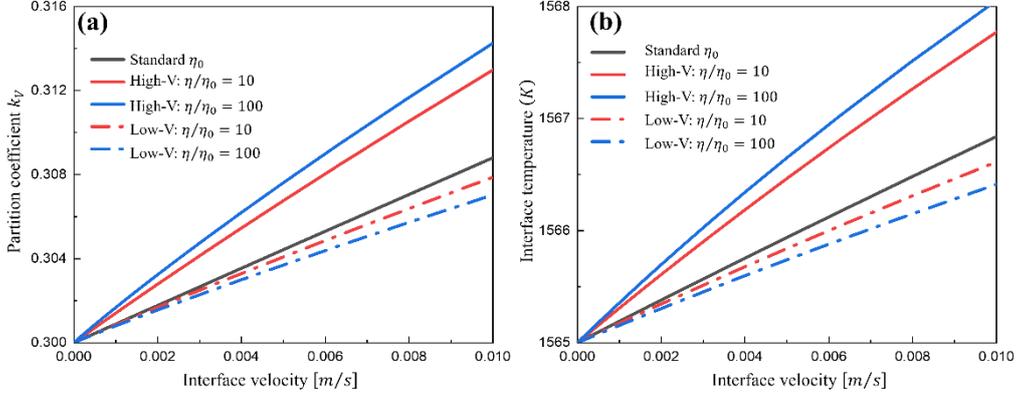

**Fig. 7.** Comparisons of 1-D steady solutions before and after enlarging the interface width from low velocity $V = 10^{-6} \, m/s$ to intermediate velocity $V = 10^{-2} \, m/s$.

Then, let us focus on $V = 10^{-6} - 10^{-2} \, m/s$, whose results are exhibited in **Fig. 7**. As theoretically expected in **Sec. 4.1**, the low-velocity solution by Eq. (20) shows better consistency with the standard solution than the high-velocity solution by Eq. (27). However, we should note that the high-velocity solution by Eq. (27) is still acceptable in this case. Alternatively speaking, although it is worse than the low-velocity solution by Eq. (20), its deviation from the standard solution is insignificant. This is different from the extension of low-velocity solution to the high-velocity regime ($V = 10^{-2} - 1 \, m/s$) in **Fig. 6**. The reason is that both physical and artificial non-equilibrium effects decrease with decreasing interface velocity, and the diffusivity enhancement by Eq. (27) is still sufficient to compensate for most artificial solute trapping. Therefore, the high-velocity solution by Eq. (27) will be finally adopted due to its better numerical stability (**Fig. 4**) and ability to represent more interface velocities (**Fig. 6** and **Fig. 7**).

Before proceeding, note that although a larger interface width $\eta/\eta_0 = 100$ has been adopted in the 1-D steady solutions, it does not mean that one can really use $\eta/\eta_0 = 100$ in practical 2-D simulation. It is because the maximum enlargement of interface width also depends on other factors like the outer diffusion length [6]. For instance, the maximum grid size $\Delta x$ should not exceed the diffusion length when matrix diffusion is significant in phase transformations. Generally speaking, the lower the interface velocity is, the larger the interface width can be adopted [6].

**5.2 Two-dimensional steady simulation**

The present Eqs. (5)-(9) and Eq. (26) constitute the governing equations for the practical numerical simulations. Herein, we use $h_s = \phi_s$, and Eq. (26) is solved to be Eq. (27) in former sections. To validate our method for practical applications, we simulate the 2-D directional solidification of the Si-9at.%As alloy, where the frozen temperature approximation is adopted, and thermal gradient is $G = 5 \times 10^6 \, K/m$. We test three pull speeds: $V_p = 0.1 \, m/s$, $0.4 \, m/s$, and $0.7 \, m/s$ and three artificial interface widths: $\eta/\eta_0 = 3, \, 5, \text{ and } 7$.

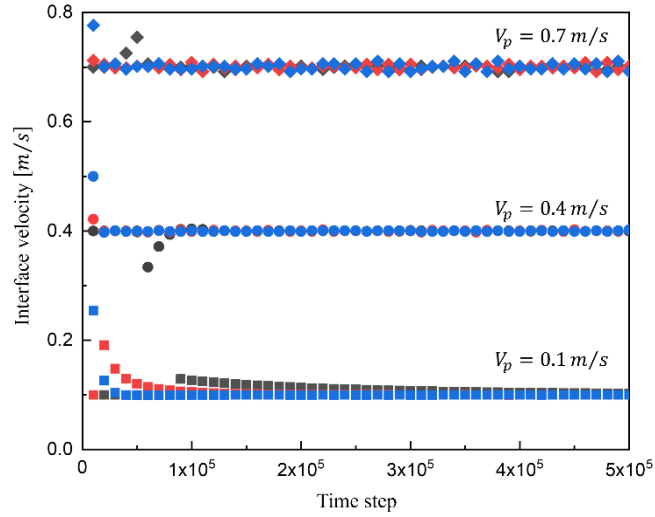

**Fig. 8.** Convergence of tip velocity for different interface width, where the black, red and blue represents the $\eta/\eta_0 = 3, \, 5, \text{ and } 7$, respectively.

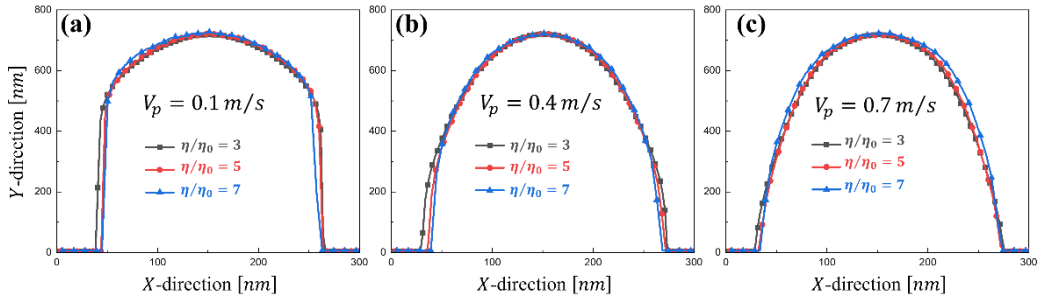

**Fig. 9.** Comparison of steady-state interface sharps for different pulling speeds and different interface widths: (a) $V_p = 0.1 \, m/s$; (b) $V_p = 0.4 \, m/s$; (c) $V_p = 0.7 \, m/s$.

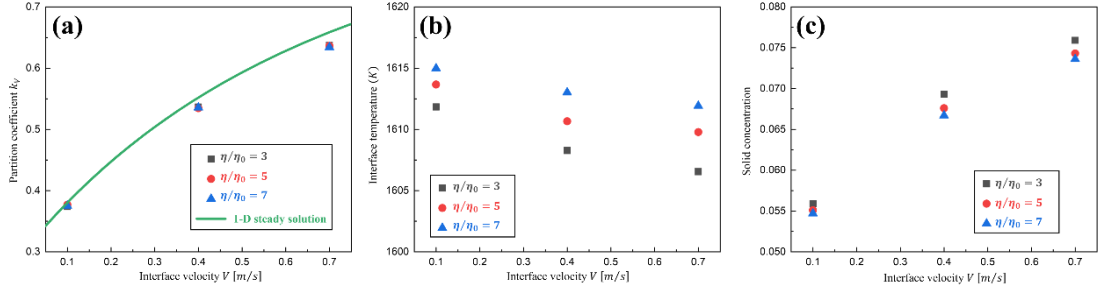

**Fig. 10.** Comparisons of steady-state (a) partition coefficient, (b) interface temperature, and (c) far-field solid concentration.

The numerical results are summarized as follows. First, results in **Fig. 8.** Convergence of tip velocity for different interface width, where the black, red and blue represents the $\eta/\eta_0 = 3$, 5, and 7, respectively. show the convergence of steady-state interface velocity. Then, comparisons in **Fig. 9** show that the steady interface sharp almost maintain after enlarging the interface width. Furthermore, results in **Fig. 10(a)-(c)** presents the convergences of steady partition coefficients, interface temperature, and far-field solid concentrations. In particular, the simulated partition coefficients are also consistent with the former 1-D steady solution since the $k_V - V$ relationship for binary dilute alloys is insensitive to the interface temperature and solid concentration. Additionally, without an overlapped point in **Fig. 10(b)**, the interface temperature seems to converge worse than the partition coefficient and solid concentration. However, the error of $\pm 5K$ is negligible compared with the absolute values of interface temperature ($\sim 1000K$). Thus, even interface temperature still converges well. Additionally, there are more obvious quantitative differences for larger $V_p$ in both **Fig. 9** and **Fig. 10**. This phenomenon was also found in Ji and Karma et al.'s work [15], and they attributed it to other artificial interface effects: surface diffusion and interface stretching. Herein, we still mainly focus mainly on suppressing artificial solute trapping, so these two effects are not fully eliminated. Even with these effects, **Fig. 9** and **Fig. 10** show that numerical simulations using the present Eq. (27) have already converged reasonably well. On the other hand, because the present method is based on the 1-D steady analysis, it is likely to be rewritten in a curvilinear coordinate and perform a more rigorous asymptotic analysis to eliminate all spurious interface effects [5,6,8,9].

## 6. Conclusions and Outlooks

Quantitative phase-field modeling has long been of interest. Despite significant progress in recent years, its description of rapid solidification is still limited to binary dilute alloys. A new quantitative phase-field framework by enhancing interface diffusivity has recently been proposed based on the idea of maintaining the physical diffuse interface kinetics. In this work, we further advance this framework by proposing a new thermodynamic and analytical method for determining the enhancement of interfacial diffusivity. The main contributions can be summarized as follows:

- Proving that phase-field kinetics are dominated by a balance of driving force and free energy dissipation, this work analytically determines interfacial diffusivity enhancement by maintaining these essential thermodynamic quantities, which can avoid inconvenient numerical corrections of partition coefficient and interface temperature.

- Based on the KKS framework and avoiding dilute approximation during model derivation, the present relationship only depends on interpolation function and artificial interface width. Thus, it is generally available for binary alloys with arbitrary thermodynamic properties.

Finally, let us conclude with a few remarks on the further extension of the present method:

- The present interfacial diffusivity enhancement can be directly applied to multicomponent alloys when diagonal atomic mobility for diffusion is dominated. Similar expressions for diffusion potential jump and free energy dissipation can be derived in such cases as those for binary alloys. The extension of considering non-diagonal mobility will be the subject of forthcoming work.

- Despite belonging to the one-side diffusion model, the present methodology can easily be extended to consider solid diffusion by adding a ratio of liquid and solid diffusivities to the present result [Eq. (26)].

- Although the present method converges reasonably well, it is still necessary to perform a more rigorous asymptotic analysis to eliminate the potential surface diffusion and interface stretching effects, which may allow a wider interface width.

**Appendix: reproducing Ji and Karma et al.'s numerical results**

In this appendix, we briefly compare our analytical method with Ji and Karma et al.'s recent numerical results [15]. Note that they used a different function for enhancing the interfacial diffusivity: $A_\phi = a - (a-1)\phi_l$. In addition, they applied the double well potential $\phi_s^2 \phi_l^2$ and the quintic interpolation function $h_l = \phi_l^3(6\phi_l^2 - 15\phi + 10)$. Substituting these functions into the present Eqs. (20) and (26), we have results in **Fig. A 1**. Herein, we can find the present high-velocity solution [Eq. (26)] agree well with Ji and Karma et al.'s numerical results by fitting partition coefficients and interface temperature thorough the entire velocity region. This also supports our finding in **Sec. 5.1**, i.e., the high-velocity solution is more suitable for enhancing interfacial diffusivity.

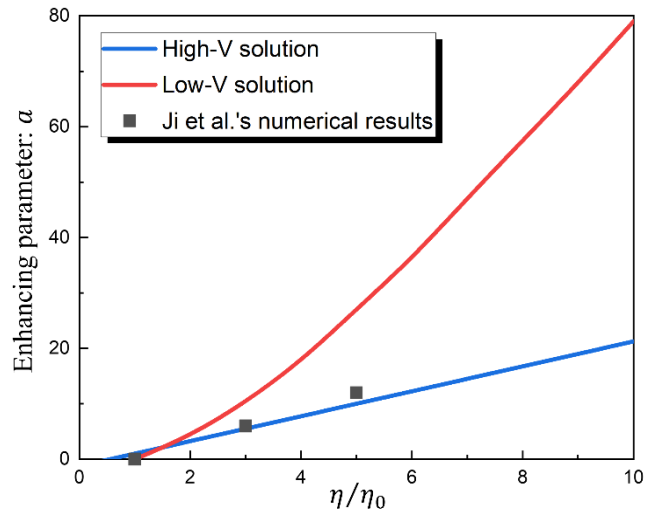

**Fig. A 1.** Comparison between Ji and Karma et al.'s numerical results with the present analytical predictions.